# Mesostructure anisotropy of bacterial cellulose—polyacrylamide hydrogels as studied by spin-echo small-angle neutron scattering

Velichko E.V.[1,2], Buyanov A.L.[3*], Chetverikov Yu.O.[2], Duif C.P.[1], Bouwman W.G.[1], Smyslov R.Yu.[2,3]

[1] Delft University of Technology, Delft, The Netherlands

[2] Petersburg Nuclear Physics Institute, National Research Center "Kurchatov Institute", Gatchina, Leningrad district, Russia.

[3] Institute of Macromolecular Compounds, Russian Academy of Sciences, Saint Petersburg, Russia

**ABSTRACT**

The submicron- and micron-scale structures of composite hydrogels based on bacterial cellulose (**BC**) and polyacrylamide were studied by spin-echo small-angle neutron scattering (**SESANS**). These hydrogels were synthesized via free-radical polymerization of acrylamide carried out in pellicle of BC swollen in the reaction solution. No neutron scattering was observed for the samples swollen in heavy water to the equilibrium state, but a SESANS signal appeared when $TbCl_3$ salt was added to the solvent. The SESANS dependences obtained for these samples revealed the anisotropy of mesostructure for the hydrogels under investigation. Density inhomogeneities on the characteristic scale of $11.5 \pm 0.5$ μm were detected in one specific orientation of the sample, i.e. with growth plane of BC parallel to plane formed by neutron beam and spin-echo length. The uniaxial anisotropy revealed agrees with the proposed model, which attributes this behavior to the existence of the tunnel-like oriented structures inside BC.

* Corresponding author: buyanov799@gmail.com

# 1. Introduction

Hydrogels are three-dimensional crosslinked structures formed by hydrophilic polymers. One of the promising fields of hydrogels' application is the development of biomaterials (implants for medical use). For example, hydrogels based on PVA are already being used as artificial cartilages to substitute the injured natural ones [1, 2]. Currently, however, the problem of improving the mechanical properties of these hydrogels and reaching characteristics of natural articular cartilage has not been completely solved [3, 4], and this situation hampers wider use of hydrogels in medicine.

In recent years, cellulose and its special type, namely bacterial cellulose (**BC**), has been used for the synthesis of various types of composite materials including hydrogel compositions for biomedical applications [5—7].

One well-known method of designing composite materials with improved functional properties is the synthesis of compositions possessing the structure of interpenetrating polymer networks (**IPNs**) [8]. We have used this method to develop the hydrogel consisting of cellulose—polyacrylamide (**PAAm**) and cellulose—polyacrylic acid (**PAA**) compounds [9-12]. Due to the high rigidity of cellulose chains, these hydrogels possess high mechanical strength and stiffness, and at the same time retain all valuable properties inherent to PAAm and PAA. These hydrogels exhibit high stiffness, strength and flexibility under different types of mechanical loads, including long-acting cyclic compression load [11, 12]. In the synthesis of our hydrogels, we used bacterial or plant cellulose as a reinforcing component.

Buyanov et al. [11] have for the first time observed an important feature of the BC-PAAm hydrogel, i.e., an anisotropy of mechanical properties, which is detected during compression. All mechanical characteristics of these materials are considerably higher when measured along the direction perpendicular to the growth surface of the original BC matrix. The authors suggested that the anisotropy of mechanical properties of BC-PAAm hydrogels is associated with structural features of BC. According to Thompson et al. [13], BC has "tunnels" oriented mainly in the vertical direction; these tunnels are formed by bacteria during biosynthesis. "Walls" of such tunnels can be condensed by congestions of the rigid chain microfibrillar BC ribbons, and these ribbons are able to mechanically reinforce hydrogels under compression in the vertical direction. In our case, tunnel lacunas in the BC structure are filled with relatively soft polyacrylamide chains, this resulting in lower compressive stiffness in the direction along the surface of the BC [11]. It should be noted that the anisotropic mechanical behavior observed experimentally can be also assigned to the existence of other types of ordering in the BC structure.

This work is aimed at studying the mesostructure of one type of BC-PAAm hydrogels by spin-echo small angle neutron scattering (**SESANS**); we intended to detect structural heterogeneities and the ordered and oriented regions. Existence of these regions is expected to lead to the observed anisotropy of mechanical characteristics. As follows from [13], the size of oriented regions in the BC structure can reach several micrometers. The use of SESANS provides information on the structural inhomogeneities with a size up to 18 or more micrometer [14], whereas the classical SANS is designed to study inhomogeneities less than a micrometer in size. SANS was previously applied to identify inhomogeneities as large as 100 nm in PAAm and PAA hydrogels [15, 16].

## 2. MATERIALS AND METHODS

### 2.1. SYNTHESIS OF BACTERIAL CELLULOSE

BC was grown by using the *Gluconacetobacter xylinus* strain (№1629 CALU) of St. Petersburg's State University (the department of microbiology) in water solutions containing 2 wt.% of glucose, 0.3 wt.% of yeast extract and 2 wt.% of ethanol at 30°C for 14 days in cylindrical glass vessels, as described in detail in [11]. The BC was subsequently washed in water solutions of potassium hydroxide at 100°C and then washed in water at room temperature. The resulting BC samples were gel-like pellicles with a thickness up to 25 mm, containing approx. 99 wt.% of water.

### 2.2. SYNTHESIS OF COMPOSITE BACTERIAL CELLULOSE—POLYACRYLAMIDE HYDROGELS

Samples of BC—PAAm hydrogels were prepared using our formerly developed technique of synthesis of composite hydrogel materials [9—12]. Acrylamide (Aldrich Chemicals) was recrystallized twice from benzene. All other reagents of analytical grade were used as received. The hydrogels were synthesized by immersing matrices containing about 1 wt.% of BC into a large amount of aqueous reaction solutions containing 55 wt.% of acrylamide for 16 h. N,N'-methylene-bis-acrylamide was used as the crosslinking agent at a concentration of $1.4 \times 10^{-3}$ M. Free-radical polymerization was initiated by cobalt (III) acetate (at a concentration of $1 \times 10^{-3}$ M); the process was conducted in cylindrical glass vessels as large as 8 cm in diameter [11]. When the synthesis was completed, the composite hydrogels in the form of a round flat layer obtained in varying thicknesses up to 3 cm were placed in distilled water for several days to remove low molecular weight components and to let the gels swell.

Equilibrium water content in hydrogels was determined gravimetrically by weighing the swollen samples and the same samples dried until constant weight was reached (at 160°C). The content of BC in BC-PAAm composition (as defined by known concentrations of BC and monomer in the reaction solution) was equal to 2 wt.% (the monomer conversion in the synthesis conditions was close to 100% [11]). The equilibrium water content in hydrogel samples was 70±3 wt.%.

## *2.3. PREPARATION OF SAMPLES FOR SPIN-ECHO SMALL ANGLE NEUTRON SCATTERING*

For SESANS experiments, rectangular blocks of $l \times w \times h$ sizes equal to 10 × 9 × 5 mm (HG1), or 16 × 10 × 5 mm (HG2) were cut out from the round flat layer of synthesized hydrogels (Table 1). To obtain luminescent hydrogels, which are available for investigating by neutron methods with contrasting original materials, the sorption of terbium chloride (III) solution was carried out in $D_2O$ by a hydrogel.

99.9% terbium chloride (III) hexahydrate ($TbCl_3 \cdot 6H_2O$) of reagent grade (CAS 13798-24-8) was purchased from Sigma-Aldrich and used without further purification to prepare aqueous ($D_2O$) solutions of two concentrations: 0.36 and 1.42 mg /mL for HG1Tb and HG2Tb, respectively (see Table 1). Sorption of $Tb^{3+}$ ions was implemented the following way: The initial block of a hydrogel (HG1 or HG2) containing $H_2O$ was first dried at 54°C until constant weight was reached. It was then placed in the 10-fold (by volume) excess of the $D_2O$ solution of terbium chloride (III) to attain equilibrium swelling at ambient temperature. It follows that the hydrogel block was as full of the $TbCl_3$ solution in "heavy" water as possible instead of "light" water.

Table 1. Hydrogels based on BC and PAAm: concentration of salt solution in $D_2O$ used for swelling ($c_{TbCl3}$), the geometry of hydrogel block after swelling

| № | Sample | $c_{TbCl3}$, mg/ml | $l \times w \times h$, mm$^3$ |
|---|--------|-------------------|-------------------------------|
| 1 | HG     | 0                 | 16×10×5                       |
| 2 | HG1Tb  | 0.36              | 16×10×5                       |
| 3 | HG2Tb  | 1.42              | 10×9×5                        |

## *2.4. METHODS*

### *2.4.1. SPIN-ECHO SMALL ANGLE NEUTRON SCATTERING*

The structure of the hydrogels was studied by SESANS. The measurements were carried out at the SESANS Delft setup (RID research reactor at the Delft University of Technology, The Netherlands) [17]. The SESANS setup has the following characteristics: wavelength $\lambda = 2.09$ Å, $\Delta\lambda/\lambda = 2\%$, range of measured spin-echo lengths z = 0.030 — 18 μm, the initial polarization of the incident beam $P = 0.93$, the sample-detector distance is fixed at 3658 mm.

The entire available range of spin-echo lengths (values of z) was employed in the measurements. The polarisation of the transmitted beam after the sample for each value of z was normalized to the one of the beam transmitted through 99% $D_2O$ to take into account the effect of the solvent, equipment characteristics, cuvettes, and background in the scattering pattern. All measurements were carried out at room temperature and atmospheric pressure. The samples consisted of rectangular blocks (Table 1) placed in quartz cuvettes with dimensions of $5 \times 10 \times 50$ mm. Each block of hydrogel with $l \times w \times h$ dimensions was measured in three orientations relative to the setup axes z and k, where z is the spin-echo length, and k is the direction of the neutron beam (Fig. 1a). Attention was paid to the orientation of the BC growth surface of synthesized HG blocks. It corresponds to the plane ($l \times w$), which is highlighted by the grid pattern in Fig. 1a.

In neutron experiments, scattering occurs due to a contrast in the neutron scattering length densities (**SLD**); in the two-phase system approach, the contrast is described by the formula:

$$\Delta\rho = |\rho_1 - \rho_2|, \quad (1)$$

where $\rho_1$ and $\rho_2$ are the SLDs of the first and second phases forming the object under study, respectively. The SLD, in turn, was calculated by the formula:

$$\rho = N\sum_i b_i = (\delta N_A/M)\sum_i b_i, \quad (2)$$

where $\delta$ is the volume density of the object, $N$ is the volume concentration of scattering centers, $M$ is the relative molecular mass, $N_A$ is the Avogadro's number, $b_i$ is the scattering length of the $i^{th}$ nucleus in the molecule. The length densities of coherent scattering $\rho_{coh}$ for the phases that constituted the objects under investigation are listed in Table 2. When calculating the contrasts, two situations were considered: homogeneous distribution of $TbCl_3$ salt between phases (then, the mass percentage in HG1Tb would be 0.024% and in HG2Tb it would be 0.105%); or selective sorption of terbium ions on BC microfibrils, then the *maximum* mass content of HG2Tb(sel) would be 1.09% assuming full absorption of terbium ions by the hydrogel from the solution.

So, physico-chemical characteristics of components of the hydrogels and the volume fraction attributed to each phase are shown in Table 2.

Table 2. Scattering length densities and phase volume fractions $\varphi_n$ of the $n^{th}$ component in the HG, HG1Tb and HG2Tb at the equal sorption of $TbCl_3$ and in HG2Tb(sel) and the selective sorption

| $n$ | Hydrogel component | Formula | $\delta^{1)}$, g/cm³ | $\rho_{coh}$ ($10^{14}$/m²) | $\varphi_n$, vol.% in hydrogel | | | |
|---|---|---|---|---|---|---|---|---|
| | | | | | HG | HG1Tb | HG2Tb | HG2Tb (sel) |
| 1 | Acrylamide unit | $(-CH_2CHCONH_2-)_n$ | 1.34 | 1.81 | 28.7 | 28.7 | 22.1 | 22.0 |
| 2 | Cellulose unit | $(-C_6H_{10}O_5-)_n$ | 1.5 | 1.76 | 0.523 | 0.523 | 0.402 | 0.401 |
| 3 | Terbium chloride | $TbCl_3$ | 4.35 | 3.56 | 0 | $6.41 \cdot 10^{-3}$ | 0.0281 | n/a$^{2)}$ |
| 4 | Hexohydrate terbium chloride | $TbCl_3 \times 6(H_2O)$ | 4.35 | 1.83 | n/a | n/a | n/a | 0.411 |
| 5 | Heavy water | D$_2$O | 1.11 | 6.34 | 70.8 | 70.8 | 77.5 | 77.2 |

Note:
[1] the density component used in the calculation of the $\rho_{coh}$ value;
[2] n/a – not applicable.

The value of transmission $T$ for hydrogels measured versus the cuvette with D$_2$O was approx. 0.7. The observed decrease of $T$ for hydrogels by ~ 30% compared to that of D$_2$O can be explained by two factors. First, the sample can contain scattering inhomogeneities, which are smaller in size than the setup resolution (< 0.030 µm). The scattering occurs at wide angles (it is known that $\theta \propto 1/d$, where $\theta$ is the scattering angle, and $d$ is the intrinsic dimension of the scattering particles), and, as a result, the scattered neutrons do not fall within the aperture before the detector. Secondly, the samples contain hydrogen atoms, and this fact leads to considerable incoherent scattering. The latter also causes the output of scattered neutrons outside the detector aperture. It is revealed that the values of $T$ for all samples are close to each other; therefore, we draw a conclusion that the above two parameters (the presence of small scale inhomogeneities and presence of incoherently scattering centers) are inherent to all samples.

In the SESANS experiment, one measures the polarization of the neutron beam as a function of spin-echo length (the distance in real space). For each spin-echo length, two measurements of the polarization are carried out: one after passing through the sample and another – after passing the setup at the same conditions, but without the sample (empty beam measurement). Then, the beam polarization after the sample is normalized to the polarization of the beam passing through the cuvette with D$_2$O. The SESANS data processing was carried out using cylindrical model (see Appendix A) [14, 18—20].

*2.4.2. PHOTOLUMINESCENCE SPECTRA*

Photoluminescence emission and excitation spectra for BC—PAAm hydrogels containing Tb$^{3+}$ ions were registered using the LS-100 BASE luminescence spectrophotometer (*PTI Lasers INC*, Canada). When using a holder for solid samples, the luminescence intensity $I_{lum}$ was recorded from the side of incidence of the exciting light beam. In the phosphorescence mode, the used integration window recorded the intensity $I_{lum}$ between 100 and 2000 µs. The grazing angle for the excitation light beam with respect to the sample was ~ 30°. Wavelength range for emission spectra was 460—700 nm at excitation wavelength $\lambda_{exc}$ = 299 nm; for excitation spectra it was 210—400 nm at the wavelength of luminescence observation $\lambda_{em}$ = 543 nm. The spectral width of monochromator slits for the excitation and luminescence was 4 nm; the PMT gain was 500. The values of $I_{lum}$ for correct comparison were reduced to an internal laboratory standard.

## 3. RESULTS AND DISCUSSION

For SESANS experiments, we have chosen one of the most interesting BC—PAAm hydrogels for practical use, which recently successfully passed the preliminary tests as artificial cartilage during *in vivo* experiments conducted on rabbits [21]. As tentative experiments showed, the measurements of the samples swollen in heavy water are not informative (no spin-echo depolarization), we have tested the option of introducing terbium ions into hydrogels by swelling the samples in a solution of $TbCl_3$ at two concentrations: 0.36 and 1.42 mg/mL for HG1Tb and HG2Tb, respectively (Table 1).

It is further known that the metal ions are capable of selective adsorption on cellulose fibrils [22, 23]. So, it is possible to assume that the sorption of terbium ions on a polymeric matrix permits a depolarization to be sufficient, especially, if this sorption might preferably occur on cellulose. The microfibrillar structure of bacterial cellulose is well studied [24]: It is formed by swollen microfibrillar ribbons (ca. 70—145 nm wide) consisting of 5 to 12 water-free $I_\alpha$-crystalline subunits with a cross-section of about 7 nm x 13 nm and of water solvating the subunits. At that, lateral aggregation of these crystalline units was found along the smaller (110)-lattice planes with a layer of water between adjacent crystallites [24]. This structure of BC can quite promote its high ability to strongly coordinated binding of metal ions, such as $Tb^{3+}$. This consideration especially holds true if it is taken according to Fink's model that there is, however, a small amount of noncrystalline tie-molecules arising from surface distortions, which connect the crystalline units laterally.

In this regard, we can consider the phase consisting of {$BC+TbCl_3 \times 6H_2O$}. Possibly, terbium ions contain aqua ligands and bound "light" water in their coordination spheres; this water can remain after the cyclic drying. Then the mass content of $TbCl_3$ salt in the hydrogel is 1.09%, i.e. an order of magnitude higher than that in the case of non-selective sorption. The volume portion of salt hydrate can reach 0.411%; this value is comparable to the portion of BC (Table 2, compare the rows 2 and 4 for the HG2Tb(sel) column).

In order to unveil the mesostructural anisotropy, all the samples were measured with SESANS in three different orientations (Fig. 1a). The dependences of the reduced polarization P (z) on the spin-echo length z for three orientations of HG2Tb (Fig. 1b, curves *1—3*) are shown in Fig. 1b, curves *1—3*). For orientations 1 and 3 (see curves *1* and *3*), the amplitude of the spin-echo polarization is close to unity throughout the entire z range studied, we can assume, therefore, absence of the inhomogeneities on this length scale. For orientation 2, a decay in P(z) is observed from 1.00 to 0.68 with further levelling-out at z values higher than 12 μm.

For HG1Tb, shapes of dependences in the orientations 1 and 3 are not shown in Fig. 1b, because they coincide with the behavior of the dependence obtained for HG2Tb. HG1Tb also demonstrates the presence of heterogeneities only in orientation 2; at the same time, the difference between the amplitudes of the spin echo signal at small and large z values (> 11 μm) is much smaller (1 to 0.9) than that for the HG2Tb sample (see curve 2', Fig. 2b). For the HG sample in all three orientations, the spin-echo signal is equal to unity throughout the whole range of spin-echo lengths, which might indicate the absence of scattering inhomogeneities on the length-scales studied.

The decrease in the amplitude of the spin-echo signal for the HG1Tb and HG2Tb samples is caused by the presence of inhomogeneities of the size lying within the z-range studied. The presence of contrast in only one of three orientations indicates a uniaxial anisotropy of the structure of the PAA–BC complexes. This anisotropy can be inherited only from the BC, which is a rigid scaffold of the entire multiphase system. The heterogeneity was observed in the direction parallel to the surface of the BC growth. The fact that a contrast was detected only in the systems containing $TbCl_3$ may be explained by both selective sorption of $Tb^{+3}$ ions on the BC microfibril surface, and by the spatial separation of PAA/BC/$D_2O$ phases in the alkaline $TbCl_3$-containing solution.

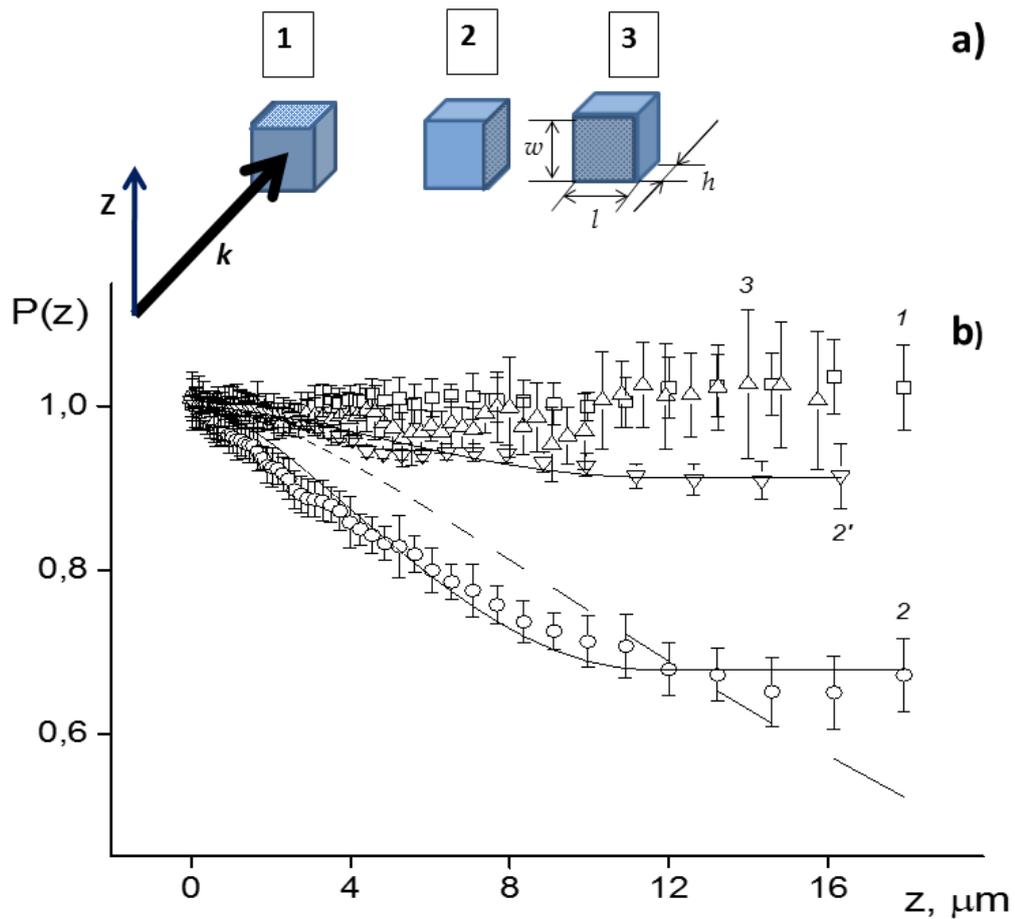

Fig. 1. The relative orientations of a block of hydrogel with dimensions $l \times w \times h$ and the BC growth surface ($l \times w$) in the neutron beam with respect to axes of SESANS setup, z and k (**a**): $z \perp (l \times w)$ *(1)*,

$(z \times k) \parallel (l \times w)$ (2), $(z \times k) \perp (l \times w)$ (3). The $(l \times w)$ surface is highlighted by a color pattern. The dependences of the reduced polarization, $P(z)$, on the spin-echo length, $z$, (**b**) for three HG2Tb orientations (1–3) and the second orientation of HG1Tb (2') are given. Solid lines present the results of fitting the experimental data using the formulas (9, 10). The dashed line corresponds to the best fit of the neutron refraction data with the formula.

To verify the validity of assumptions about selective sorption, this mechanism was investigated by fluorescent methods. Fig. 2 and 3 show the excitation and emission spectra of the luminescence hydrogels exposed to $TbCl_3$ solution in $D_2O$ until equilibrium swelling was reached. The luminescence spectrum of HG1Tb (Fig. 2a) has a quasi-lined structure with characteristic peaks in the luminescence band at 490, 543, 580, 620 nm etc., which can be compared with the atomic resonance transitions in the luminescence spectra of $Tb^{3+}$ ions, corresponding to the spectral lines $^5D_4 - {}^7F_J$, where J = 6—0 [25]. The same pattern was found in the case of the HG2Tb spectrum.

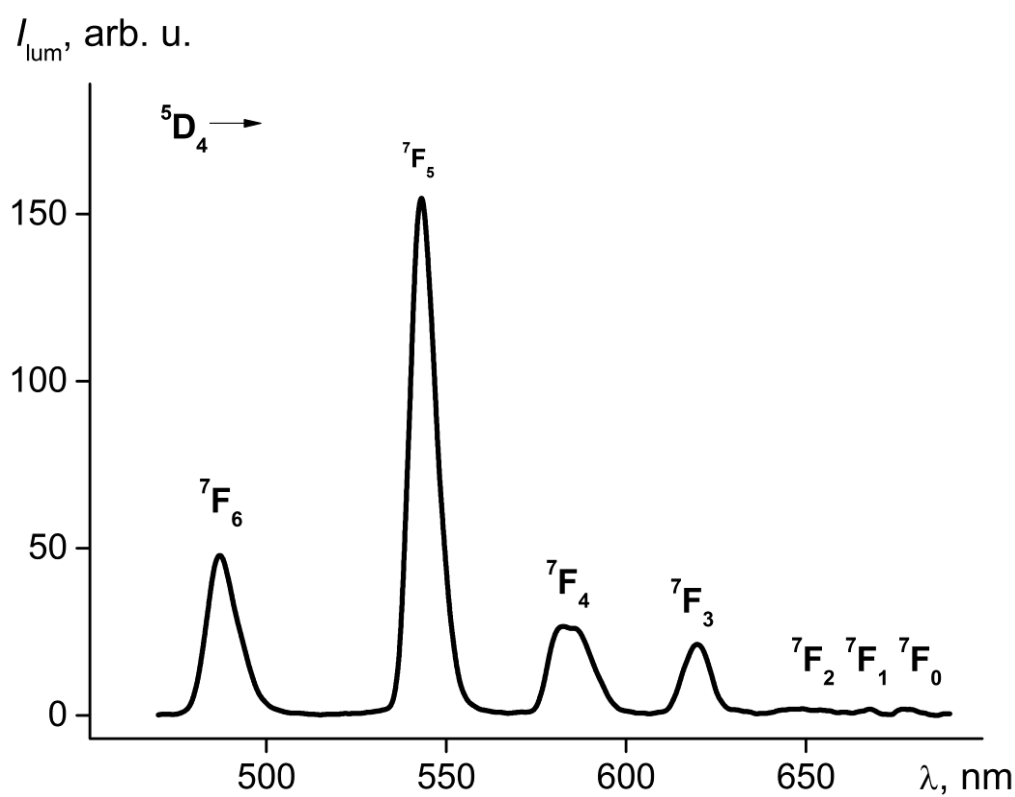

Fig. 2a. Luminescence emission spectrum for HG1Tb. Luminescence excitation was carried out at 299 nm.

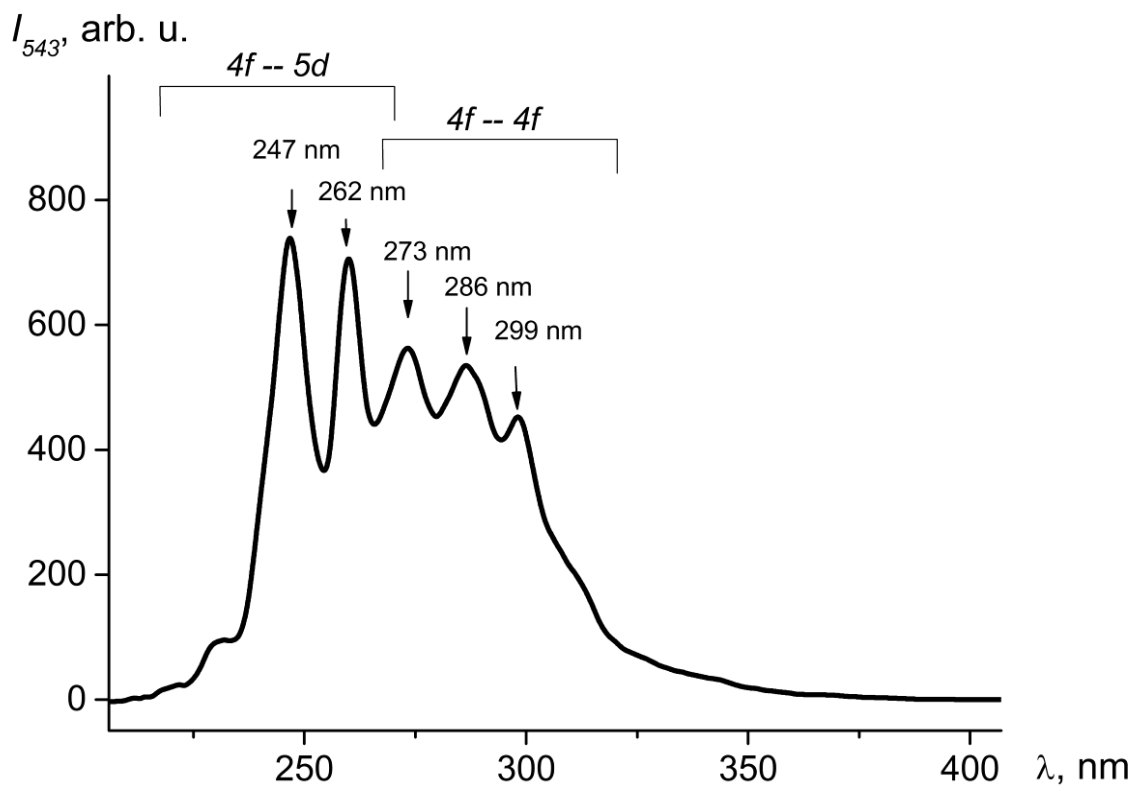

Fig. 2b. Luminescence excitation spectrum for HG1Tb. Luminescence was observed at 543 nm.

In the luminescence excitation spectrum for HG1Tb (Fig. 2b), peaks at 247 and 262 nm corresponding to *4f—5d*-transitions are present, and the peaks at 273, 286, and 299 nm are attributed to *4f—4f* transitions [26, 27]. For TbCl$_3$ solution in "heavy" water, these intense contributions at the above wavelengths were not observed. The same pattern was found for HG2Tb.

The resulting increase in the efficiency of luminescence excitation for HG1Tb and HG2Tb may be due to the fact that Tb$^{3+}$ ions provide their virtual *5d* and *4f* atomic orbitals for lone electron pairs of the hydroxyl oxygen atoms of water and the matrix. Thus, it is possible to form a donor-acceptor coordination bond between Tb$^{3+}$ ions and the matrix of the hydrogel. It is important that the bound organic compounds located near the central atom of the complex can absorb with a larger extinction coefficient in the range of 280—310 nm than lanthanide ions themselves [26]. Then, due to the resonant dipole–dipole interaction, electronic excitation energy is transferred from the "antenna" of the organic matrix to an isoenergetic level of Tb atom according to the Förster resonance mechanism (FRET). It can be concluded that Tb$^{3+}$ ions are adsorbed by the hydrogel matrix; i.e., we are talking about a possible mechanism of selective sorption of Tb$^{3+}$ ions on cellulose microfibrils.

The data obtained (by SESANS, luminescence) shows that it is impossible to account formally for the volume fraction (portion) of water in a biphasic approximation. After all, the values of $\varphi_n$ in HG2Tb and HG2Tb(sel) are practically the same (Table 2). However, the value $P(z)$ for the 2$^{nd}$ orientation decreases depending on the concentration of $Tb^{3+}$. The conclusion might be made that the water in the complex in the form of aqua ligands, the bound water, and the water just in the lacunas may produce different contributions when calculating the contrast [28]. For example, it is seen that for solutions of protein, surrounding hydration layer by its contrast differs from both the protein and a bulk solvent [29, 30].

Strong uniaxial anisotropy is described by the model of oriented cylinders. It can be explained by the presence of "tunnels" (hollow cylinders) in the sample with walls formed from compacted bacterial cellulose; these tunnels are filled with PAAm. The main volume of the sample outside the channels is also filled with PAAm.

Fig. 1b (curves *2* and *2'* for HG2Tb and HG1Tb, respectively) presents the results of fitting the experimental data with the use of Eq. (12).

Fitting yielded the following parameters: $D = (11.5\pm0.5)$ μm, $\varphi = 0.05\pm0.01$ for both samples; the contrast ($\Delta\rho_0$) for the HG1Tb sample is $0.3\times10^{14}$ m$^{-2}$, and for the HG2Tb, the value was $0.49\times10^{14}$ m$^{-2}$. The value of D does not depend on the above assumption about selective sorption.

Using the concept of curved tubes, one can make the assumption that the channels are oriented perpendicularly to the BC growth surface, and their length substantially exceeds 20 μm (the maximum size currently available in the SESANS method). Under these conditions, it is easy to see that in the sample orientation 1, the experimental signal should be absent according to Eq. (10), since L→∞. For ideally oriented cylinders, the signal from the sample in orientation 3 should be observed as well as in the case of direction 2 (see Fig. 2A). However, the signal in orientation 3 was not observed: It leads to our doubts about the correctness of the used model. Another possible reason that we obtained a SESANS signal in only one orientation of the sample is neutron refraction. In this case, we should observe a signal in only one orientation of the sample, i.e. with the cylinders perpendicular to both the spin-echo length and neuron beam axes. The dashed line in Fig. 1b represents the best fit obtained for curve 2 with assuming neutron refraction. As can be seen, this fit does not describe the obtained signal. Therefore, none of the models used seem to describe the obtained data completely, but we assume that neutron refraction might cause an increase of the SESANS signal for oriented cylinders in orientation 2, which might lead to observed results. Thus, the only registered signal in orientation 2 can be described by the correlation function of a cylinder with the long side oriented along the y-axis, and the observed dependence suggests that the diameter of the cylinder lies in the range of 11.5±0.5 μm.

## 4. Conclusions

1. According to the SESANS study, the BC—PAAm hydrogels under investigation have been proven to possess a uniaxial anisotropy of micron structure with a characteristic size of 11.5±5 μm, the axis being normal to the BC growth surface. The contrast of the inhomogeneities of the structure manifested itself when $TbCl_3$ salt was added to the solvent.

2. The excitation spectra have revealed the intensive photoluminescence of hydrogels containing $Tb^{3+}$ ions in the region from 280 until 310 nm. It indicates that these ions from solution are associated with the hydrogel matrix.

3. Thus, the results obtained by SESANS confirmed the assumption made in [11] that the anisotropy of the mechanical properties of these systems is caused by the specific structure of the BC matrix in the hydrogels which, according Thompson et al. [13], has the oriented tunnel-like organization.

**ACKNOWLEDGEMENTS**

The authors are grateful to Dr. A.A. Tkachenko and Dr. A.K. Khripunov for providing samples of the bacterial cellulose. The authors are also thankful to Reactor Institute Delft for providing the beam time. The authors kindly acknowledge the financial support of the Presidium of the Russian Academy of Sciences (RAS) (the grant FIMT-2014-066). Dr. Yu.O. Chetverikov appreciates Russian Foundation for Basic Research (grant № 16-02-00987) for financial support.

**APPENDIX A. SUPPLEMENTARY DATA**

Analytical dependence of the reduced polarization on the spin-echo length is described by the following equation [14]:

$$P(z) = \frac{P_{sample}(z)}{P_{solvent}(z)} = e^{\Sigma_t(G(z)-1)}, \qquad (3)$$

$$\Sigma_t = t\lambda^2(\Delta\rho)^2\varphi(1-\varphi)\xi, \qquad (4)$$

where $z$ is the spin-echo length (varied in the SESANS experiments); $t$ is the thickness of the sample; $\Sigma_t$ is the average number of scattering events of the neutron in the sample; $G(z)$ is the projection of the pair correlation function on the axis codirectional with z; $\varphi$ is the volume fraction of any of the phases in the two-phase system, $\xi$ is the correlation length of the scattering inhomogeneities; λ is the neutron wavelength.

The dependence of $G(z)$ is a projection of the spatial correlation function $\gamma(r)$ on the quantization axis, along which correlations are measured in SESANS (spin-echo length, z):

$$G(z) = \int_z^{+\infty} \frac{\gamma(r)}{\sqrt{r^2 - z^2}} dr, \tag{5}$$

where the spatial correlation function γ(r) is defined as

$$\boldsymbol{\gamma(r)} = \int_V \boldsymbol{\rho(r)\rho(r+r')dr'}, \tag{6}$$

where the integration is performed over the volume V of the system using the radius vector *r* = (*x, y, z*) and *r'* = (*x', y', z'*).

From the experimental point of view, the *G(z)* function is retrieved by

$$G(z) = \frac{\ln P(z)}{\ln P(\infty)}, \tag{7}$$

where *P*(z) is the polarization of the neutron beam which passed through the sample and normalized to the polarization after passing through the cuvettes with solvent. The *P*(∞) value characterizes the fraction of the neutrons that are not scattered during transmission through the sample. Moreover, the value of the saturation level also provides information about the concentration by taking ϕ into account, information about the microstructure via *ξ*, and about the chemical structure through Δ*ρ* [14]:

$$P(\infty) \equiv e^{-\Sigma t}. \tag{8}$$

As was mentioned before, due to strong anisotropy of the mechanical characteristics of the samples under investigation, it was expected to observe an orientation-dependent SESANS signal. Therefore, each sample was measured in three orthogonal orientations, schematically represented in fig. 1a. In case of an anisotropic sample structure, it is likely to observe different SESANS signals for different sample orientations relatively to the setup geometry [19]. In the present work, two different ideas were considered to describe the obtained experimental results: a scattering model of oriented cylinders and neutron refraction [20].

In case of oriented cylinders we will consider cylinders oriented along three different axes of Cartesian coordinates (x, y, z), connected with experimental setup. X-axis corresponds to the neutron beam direction; the z-axis is the vertical axis along which the spin-echo length is being varied in the experiment; the y-axis is perpendicular to the former two and it should be noted that SESANS is insensitive to inhomogeneities in the y-direction. A detailed derivation of *G(z)* functions for all cylinder orientations might be found in [14], here we will only show the G(z) functions and corresponding correlation lengths expressions. As a starting point, we shall consider the autocorrelation function of a disc in 2 dimensions:

$$\gamma_c(r) = \begin{cases} \frac{2}{\pi} \left( \cos^{-1}\left(\frac{r}{D}\right) - r \frac{\sqrt{D^2 - r^2}}{D^2} \right) & \text{if } r \leq D \\ 0 & \text{otherwise} \end{cases} \tag{9}$$

For a cylinder with the long side coaxial with the z-axis of the setup and the cross section in the xy-plane we will have:

$$G(z) = 1 - z/L. \tag{10}$$

By integrating the former function over the diameter D, we will get the correlation length:

$$\xi = 2\int_0^D \gamma_c(x)dx = \frac{8D}{3\pi} \tag{11}$$

which is the average length of all chords drawn inside a disk of diameter D.

Having the side of the cylinder parallel to y and its diameter in the *xz* plane yields, for z< D:

$$G(z) = \int_0^D \gamma_c\left(\sqrt{x^2 + z^2}\right)dx, \tag{12}$$

and again the correlation length is $\xi = 8D/3\pi$.

If the cylinder is oriented with its side along x and its face in the zy plane, the projection along x becomes, for z< D:

$$G(z) = \gamma_c(z) \tag{13}$$

and the correlation length is in this case L:

$$\xi = 2\int_0^D \left(1 - \frac{x}{L}\right)dx = L. \tag{14}$$

In case of neutron refraction, the polarization of the neutron beam exponentially depends on the number of refractive elements on the beam pathway:

$$P(z) = (2\Delta\rho\lambda z \cdot K_1(2\Delta\rho\lambda z))^n, \tag{15}$$

where $K_1(x)$ is the first-order modified Bessel function of the second kind. In case of oriented cylinder-like refractive elements, we can expect this effect only in the orientation of cylinders perpendicular to both spin-echo length and beam axes.